# Control of planar nonlinear guided waves and spatial solitons with a left-handed medium

A. D. Boardman, P. Egan, L. Velasco and N. King

Joule Physics Laboratory, Institute for Materials Research, University of Salford, Salford, Manchester, M5  4WT, United Kingdom

a.d.boardman@salford.ac.uk
http://www.imr.salford.ac.uk/groups/photonics/index.shtml

## Abstract

The evidence that double negative media, with an effective negative permittivity, and an effective negative permeability, can be manufactured to operate at frequencies ranging from microwave to optical is ushering in a new era of metamaterials. They are referred to here as 'left-handed', even though a variety of names is evident from the literature. In anticipation of a demand for highly structured integrated practical waveguides, this paper addresses the impact of this type of medium upon waveguides that can be also nonlinear. After an interesting historical overview and an exposure of some straightforward concepts, a planar guide is investigated, in which the waveguide is a slab consisting of a left-handed medium sandwiched between a substrate and cladding that are simple dielectrics. The substrate and cladding display a Kerr-type nonlinear response.  Because of the nonlinear properties of the Kerr media, the power flow direction can be controlled by the intensity of the electric field. A comprehensive finite-difference-time-domain (FDTD) analysis is presented that concentrates upon spatial soliton behaviour. An interesting soliton-lens arrangement is investigated that lends itself to a novel cancellation effect.

## 1. Introduction

The interaction between polarisation excitations of condensed matter and electromagnetic waves still attracts a lot of attention. This is because an electromagnetic wave travelling through a polarisable medium is modified by the polarisation it induces, and becomes coupled to it. A typical host is an electron plasma, which is an adequate description of a metal at wave frequencies that do not require the compensating lattice to be brought into the model. If the relative permittivity function for the electron gas, in the frequency domain, is a scalar function $\varepsilon(\omega)$, for an angular frequency $\omega$, then bound non-radiative surface modes [1] can be associated with the regime $\varepsilon(\omega) < 0$.  Specifically, a resonance occurs when $\varepsilon(\omega) = -1$ , and  $\omega = \omega_p / \sqrt{2}$ . Here $\omega_p$ is the bulk plasma frequency and the relative permeability, in this case, is simply $\mu = 1$. Over several decades, magnetic materials, characterised by a frequency-dependent relative permeability tensor function **μ**($\omega$), have also attracted a lot of attention, with ferrimagnets [2] being the principal target materials for investigation.  For many magnetic materials, however, dielectric anisotropy can be ignored and the dielectric constant is usually just a scalar constant, often set equal to unity. Unfortunately, the familiar treatment of Maxwell's equations that introduces polarisation and magnetisation is ambiguous [3], so it is not clear that a material description in terms of *both* a relative permittivity *and* a relative permeability has a clear meaning. This is especially true at optical frequencies [4] at which is difficult to sustain a proper interpretation of relative permeability.  Indeed, it has been suggested that the familiar separation, in which current density and charge in Maxwell's equations are abandoned in favour of polarisation and magnetisation vectors be given up in favour of an all-embracing, effective, relative permittivity tensor. The interesting thing about this approach is its impact upon the widely used boundary conditions. If the continuity of tangential components of magnetic field **H** and electric field **E** and the normal components of magnetic flux vector **B** and displacement vector **D** are to be retained, then the separation in terms of relative permittivity and permeability tensors must be introduced. If an effective permittivity tensor is introduced then the boundary conditions become rather complicated [3], so the treatment given here will assume that an accurate description is given by maintaining constitutive relationships that introduce *both* permittivity and permeability.  In any case, it has been shown recently [5] that, in fact, both descriptions correspond precisely with each other and so it is completely acceptable to adopt such a separation.

In magneto-optics [6], the $\varepsilon(\omega)$ tensor depends upon the magnetisation of the material and $\mu = 1$, but in a study of gyroelectromagnetics [7], and also the interaction of spin, plasma and electromagnetic waves [8], both $\boldsymbol{\varepsilon}(\omega)$ *and* $\boldsymbol{\mu}(\omega)$ are in operation. It is not unusual then to consider both permittivity and permeability in the same material, but in general these tensors have off-diagonal elements and have properties that are very specific with respect to the applied magnetic field orientations. In recent years, a lot of excitement has been generated about *isotropic* materials that are characterised by the *scalar* functions $\varepsilon(\omega)$ *and* $\mu(\omega)$. They have very unusual properties when $\varepsilon(\omega) < 0$ *and* $\mu(\omega) < 0$ simultaneously [9], not least being their behaviour when a frequency can be found that pins down the condition $\varepsilon(\omega) = -1$ *and* $\mu(\omega) = -1$. The latter is the basis for proposals to use plane-parallel lenses [9-12] for aberration-free imaging. It is also possible to engage in what is now termed *negative refraction* with such materials [11-14]. The creation of this type of refraction relies, in this case, upon the conditions $\varepsilon(\omega) < 0$ *and* $\mu(\omega) < 0$, and does not need anything other than an isotropic substance. The creation of these materials at optical frequencies is gaining momentum, with an impressive measure of success [15,16] being in evidence from metal nanowire composite materials. The latter hold out the exciting promise to bring the $\varepsilon(\omega) < 0$ *and* $\mu(\omega) < 0$ condition into the optical domain. They complement very nicely the pioneering experiments at microwave frequencies that have provided such a huge stimulus to the field [13,14].

An important, and critical, physical consequence of the $\varepsilon(\omega) < 0$ *and* $\mu(\omega) < 0$ condition is the appearance of *backward* waves [17] in the material. This is a very simple consequence that flows from a simplistic plane wave treatment of Maxwell's equations, after endowing them with a separation of the current density into terms involving polarisation and magnetisation, in the manner discussed earlier [3]. Taking the two Maxwell curl equations, and assuming that a plane wave $\exp i(\omega t - \mathbf{k}.\mathbf{r})$ is propagating, where $\mathbf{k}$ is the wave vector, $\mathbf{r}(x, y, z)$ is the spatial vector involving the coordinates $(x, y, z)$, and $t$ is the time coordinate, produces

$$\mathbf{k} \wedge \mathbf{E}(\omega) = \omega\mu_0\mu(\omega)\mathbf{H}(\omega)$$
$$\mathbf{k} \wedge \mathbf{H}(\omega) = -\omega\varepsilon_0\varepsilon(\omega)\mathbf{E}(\omega) \tag{1}$$

where $\mathbf{E}(\omega)$ and $\mathbf{H}(\omega)$ are complex and are the Fourier transforms of the field vectors. It is now expedient to adopt the elementary triple vector products

$$\mathbf{E} \times (\mathbf{k} \times \mathbf{E}^*) = \mathbf{k}(\mathbf{E} \cdot \mathbf{E}^*) - \mathbf{E}^*(\mathbf{E} \cdot \mathbf{E}) = \mathbf{k}|\mathbf{E}|^2$$
$$(\mathbf{k} \times \mathbf{H}) \times \mathbf{H}^* = -\mathbf{H}^* \times (\mathbf{k} \times \mathbf{H}) = -[(\mathbf{H} \cdot \mathbf{H}^*)\mathbf{k} - (\mathbf{H}^* \cdot \mathbf{k})\mathbf{H}] = -\mathbf{k}|\mathbf{H}|^2 \tag{2}$$

These imply that the Poynting vector $\mathbf{S}$ is

$$\mathbf{S} = \frac{1}{2}(\frac{1}{\omega\mu_0\mu(\omega)})\mathbf{k}|\mathbf{E}|^2 = \frac{1}{2}(\frac{1}{\omega\mu_0\varepsilon(\omega)})\mathbf{k}|\mathbf{H}|^2 \tag{3}$$

It is immediately apparent that $\mathbf{k}$ is *anti-parallel* to $\mathbf{S}$, under the condition $\mu(\omega) < 0$ and $\varepsilon(\omega) < 0$ and that *backward waves* can be expected. At this stage of the argument, nothing has been assumed about the refractive index and it can be set to the positive value $n = \sqrt{\mu(\omega)\varepsilon(\omega)}$, if that is desired. In fact $n$ can be set to a positive *or* negative value: the choice does not affect the necessary appearance of backward waves but it is derived from the fact that any square root offers the options ±. The vectors $(\mathbf{k}, \mathbf{E}(\omega), \mathbf{H}(\omega))$ now form a *left-handed* set : a term that comes from the reference [18] cited in the early paper by Veselago [9]. Although it now seems obvious, with hindsight, a general excitement concerning the use of $\mu(\omega) < 0$ and $\varepsilon(\omega) < 0$ materials, under the heading left-handed materials is very much under way.

The possibility that backward waves can occur in isotropic optical media was first discussed by Schuster, having been stimulated by the work of his Manchester colleague Lamb, the great hydrodynamicist [19]. Schuster [20] stated that one curious result follows: *the deviation of a wave entering such a medium is greater than the angle of incidence so the wave normal is bent over to the other side of the normal [sic]*. In other words, Schuster understood very clearly that light upon entering such a medium "bends the wrong way" and that *negative refraction* occurs. He also discussed the possibility of observing what he called convection of light energy forward with waves moving backwards [sic] and pointed out that, although there are optical media where dispersion is reversed, it is in regions of extremely high absorption, and so doubted whether the phenomenon had any application. Schuster, by drawing attention to regions of high optical absorption in which dispersion is reversed[sic] did not rule out negative refraction; he was just pessimistic about its applications He cited strong solutions of fuchsin or cyanin, and was less than enthusiastic about the term *anomalous dispersion*, preferring, instead, to call the phenomenon *selective refraction*. This type of negative refraction involving backward waves does fall into the left-handed class but this type of refraction can occur in a much broader sense without having to insist upon the condition $\mu(\omega) < 0$ and $\varepsilon(\omega) < 0$. Later on it was called negative dispersion and the refractive consequences were investigated long ago by Mandelshtam [21,22]. Today the term negative refraction has gained common currency, and the race is on to find metamaterials that display it. $\mu(\omega) < 0$ and $\varepsilon(\omega) < 0$ materials look as if they may are suitable for modern applications in the microwave and optical frequency ranges [23] but it is important to note that negative refraction will occur more widely. For example, the study of particle accelerators and microwave electron tubes contains discussion about the fact that periodic guiding structures always supports backward waves [24], or what that field continues to call anomalous dispersion [25]. For this reason photonic crystals [26] have also revealed the possibility of negative refraction. Another classic example involves uniaxial crystals. Negative refraction will occur even with a $45^0$-cut calcite crystal [27,28] in air [sic] but the group velocity and the wave velocity are not antiparallel in this case but are set at an angle to each other. This is an effect that is entirely due to the crystal anisotropy and it is not what is investigated in this paper. Crystal anisotropy generates an easily observable example of negative refraction that, while it has its own importance is part of a broad church containing negative refraction phenomena. The emphasis in left-handed work is to abandon anisotropy in favour of isotropy and set the triad of vectors (**E**, **H**, **k**) to form a *left-handed set*. This is what is fuelling the search for new metamaterials. Future demand will be satisfied by highly structured metamaterials that can be used in a practical way at frequencies ranging all the way up to the optical bands. Nonlinearity will also be a companion to this quest and this is addressed here. Finally, losses are considered to be negligible throughout this study but, although it is acknowledged that real materials will bring in losses, the general principles exposed here will still be viable. The underlying condition may have to altered to acknowledgement the existence of complex relative permeability and permittivity $\mu = \mu_r + i\mu_i, \varepsilon = \varepsilon_r + i\varepsilon_i$. This will imply that [29] that $\mu_r(\omega) < 0$ *and* $\varepsilon_r(\omega) < 0$ simultaneously has the status of a sufficient condition, rather than the a necessary one implied in the earlier part of the discussion. The conclusions to be drawn here need to be experimentally demonstrated in regimes that minimise losses, in order to preserve the qualitative properties. This can be done through a variety of mechanisms, including nonlinear gain but they will be addressed in a later publication.

## 2. Surface waves of a nonlinear left-handed planar thin film structure

The existence of linear surface polaritons at a planar interface between a semi-infinite plasma and a dielectric has been known for many decades. However, since the electric field amplitudes must approach zero as the distance from the boundary approaches infinity, only TM surface waves can exist. This can be viewed as a consequence of the boundary conditions since, for TM waves, the electric field is continuous at the boundary whereas the gradient of the electric field has the opposite sign at each side of the boundary. This means that for TM surface waves a field profile that decreases on each side of the boundary satisfies the dispersion equation. On the other hand, for TE surface waves, both the electric field and its gradient at the boundary must be continuous. This precludes the possibility of a decrease in the field at each side of the boundary which is a necessary condition for the existence of TE surface waves.

A model structure that produces TE surface waves at a single boundary can be demonstrated theoretically [30]. The necessary condition of continuity of both the electric field and its derivative is satisfied if one of the semi-infinite media displays a positive Kerr type nonlinearity. Maxwell's equations then generate a *sech* type solution in the nonlinear medium so that, under the influence of the electric field intensity, the field reaches a maximum and then decreases, approaching zero towards infinity. However, large field intensities are necessary for this to occur.

More recently [31], it has been shown, again theoretically, that if one of the semi-infinite media is a *left-handed medium* (LHM) then the boundary conditions for TE waves requires that the gradients of the electric fields on each side of the boundary must now have opposite signs. This means that TE surface polaritons can exist without the need for large electric field intensities. The conclusions are based upon the assumption that the electromagnetic properties of the left-handed medium can be separated into a scalar effective relative permittivity $\varepsilon(\omega) = \varepsilon_{eff}(\omega)$ and scalar effective relative permeability $\mu(\omega) = \mu_{eff}(\omega)$, where.

$$\varepsilon_{eff}(\omega) = 1 - \frac{1}{\tilde{\omega}}; \quad \mu_{eff}(\omega) = 1 - \frac{F\tilde{\omega}^2}{\tilde{\omega}^2 - \tilde{\omega}_0^2} \tag{4}$$

Here the normalised frequency is, $\omega_p$ is an effective plasma frequency that depends upon the geometry of the system, $\omega_0$ is a resonance frequency and $F$ is parameter that also depends upon the system structure. Using these forms for $\varepsilon_{eff}(\omega)$ and , $\mu_{eff}(\omega)$, TE surface modes can be generated with a dispersion curve has a negative gradient, indicating a *negative group velocity*. This negative group velocity depends critically on the permittivity of the right-handed medium (RHM), and that above a critical value $\varepsilon_c$, given by

$$\varepsilon_c = \frac{2-F}{2\tilde{\omega}^2} - 1 \tag{5}$$

the gradient of the dispersion curve is positive. Hence, for permittivities greater than the critical value, the group velocity is positive, as shown figure 1.

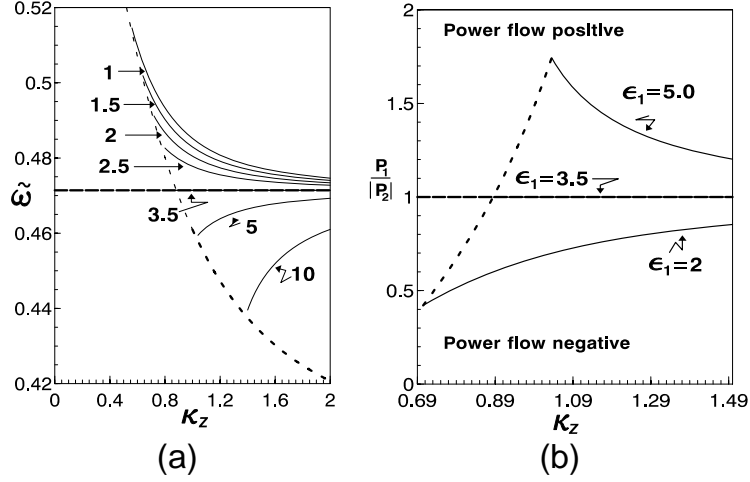

(a) (b)

Figure 1 (a) *The dispersion curves for a single interface between a dielectric and a LHM. Curves are labelled with the permittivity of the dielectric and show the critical value of the permittivity where the group velocity goes from positive to negative. The values for $\omega_p$, $\omega_0$ and F are those given in reference 1. $\tilde{\omega} = \omega/\omega_p$ $\kappa_z = ck_x/\omega_p$. At the critical value $\omega = \omega_0\sqrt{\dfrac{2}{2-F}}$. The dashed curve is the limit of the surface wave region where $k_z = 0$, (b) The power flow for typical curves from* (a).

The boundary between a semi-infinite LHM and a semi-infinite Kerr type nonlinear medium[32] also supports TE surface waves, and that the critical value (5) that is evident in the linear case has a counterpart in the nonlinear/LHM case, where the permittivity of the RH nonlinear medium is a function of the electric field intensity. For a zero-field permittivity $\varepsilon_1$, this critical value of electric field is

$$E_c^2 = \frac{\omega_p^2(2-F)}{2\omega_0^2} - (1-\varepsilon_1) \qquad (6)$$

where $E_c$ is the amplitude of the field at the boundary. Thus, whether the power flow is in the positive, or negative, direction is determined by the value of the amplitude on the boundary compared with the critical value $E_c$.

Similar characteristic behaviour is found here for a planar waveguide bounded by a substrate and a cladding either, or both, of which display a positive Kerr-type nonlinearity. The *linear* dispersion equation for TE surface modes of the form $Be^{k_f x + i(k_z z - \omega t)} + Ce^{-k_f x + i(k_z z - \omega t)}$ is

$$\tanh(k_f d) = -\frac{k_f \mu_f (\mu_s \alpha_c + \mu_c \alpha_s)}{(k_f^2 \mu_s \mu_c + \alpha_s \alpha_c \mu_f^2)} \qquad (7)$$

for a film ($0 < x < d$), and fields of the form $Ae^{\alpha_s x + i(k_z z - \omega t)}$ in the substrate ($x < 0$), and $De^{-\alpha_s(x-d) + i(k_z z - \omega t)}$ in the cladding ($x > 0$). The propagation direction is parallel to z- axis and $\mu_f$, $\mu_s$, $\mu_c$ are the permeabilities of the film, the substrate and the cladding, respectively. The transverse wave number is $k_f = \left( k_z^2 - \frac{\omega^2}{c^2} \mu_f(\omega) \varepsilon_f(\omega) \right)^{1/2}$. The negative sign on the right-hand-side of (7) negates the possibility of a solution and therefore the existence of TE surface waves in this model. However, it was shown by Boardman and Egan [33] that if either one, or both, of the bounding dielectrics is nonlinear, then surface TE waves *can* exist for high intensity electric fields where, as in the single interface case, there is a *sech* type solution to Maxwell's equations in the substrate and cladding. Now examination of (7) shows that if the film is a LHM then there is also a solution for TE surface waves when the substrate and cladding are linear [34].

In more detail, it is interesting to investigate the case in which the film is a LHM *and* the substrate and/or cladding are nonlinear dielectric right-handed media, displaying a Kerr type nonlinearity. To be specific, a nonlinear substrate and linear air cladding will be used for the numerical calculations..

For TE waves in the non-linear boundary layers, where $\mu_{s,c} = 1$, $\gamma_{s,c}$ are constant nonlinear factors, $c^2 = 1/\varepsilon_0 \mu_0$, $\varepsilon_{s,c}$ are the relative permittivities of the substrate and cladding, respectively, and there is no variation in the *y*-direction, Maxwell's equations, for the substrate and cladding reduce to

$$\frac{\partial^2 E_{s,c}}{\partial x^2} - \left( k_z^2 - \frac{\omega^2}{c^2} \varepsilon_{s,c} - \gamma_{s,c} \frac{\omega^2}{c^2} |E_{s,c}|^2 \right) E_{s,c} = 0 \tag{8}$$

where $E_{s,c}$ are the amplitudes of the electric field in the substrate and cladding respectively. We now assume that $E_{s,c}$ are real (ref 1) so that integrating (5) gives

$$\left( E'_{s,c} \right)^2 - \left( k_z^2 - \frac{\omega^2}{c^2} \varepsilon_{s,c} - \frac{\omega^2}{c^2} \frac{\gamma_{s,c}}{2} E_{s,c}^2 \right) E_{s,c}^2 = K \tag{9}$$

where the prime represents a derivative with respect to *x* and *K* is a constant of integration. Using the field in the film, like the linear boundary case above, together with (8) gives the four dispersion equations for an LHM film bounded by semi-infinite nonlinear substrate and cladding, i.e.

$$\tanh(k_f d) = \pm \frac{k_f \mu_f \left[ \sqrt{k_z^2 - \frac{\omega^2}{c^2} \left( \varepsilon_c + \frac{\gamma_c}{2} E_b^2 \right)} \pm \sqrt{k_z^2 - \frac{\omega^2}{c^2} \left( \varepsilon_s + \frac{\gamma_s}{2} E_0^2 \right)} \right]}{k_f^2 \pm \mu_f^2 \sqrt{\left[ k_z^2 - \frac{\omega^2}{c^2} \left( \varepsilon_c + \frac{\gamma_c}{2} E_b^2 \right) \right] \left[ k_z^2 - \frac{\omega^2}{c^2} \left( \varepsilon_s + \frac{\gamma_c}{2} E_0^2 \right) \right]}} \tag{10}$$

where it is understood that the signs in the numerator and the denominator on the right-hand-side are the same. Thus there are four dispersion equations for the nonlinear boundary layer model and the choice of signs will determine the field profiles in the boundary regions. The choice of a minus sign before the left-hand-side of (10) and minus signs in the numerator and denominator corresponds to an electric field profile that decays immediately away from the cladding boundary but increases, in the first instance, into the substrate, as seen in figure 2. It can be seen from this figure that very high-intensity electric fields would be necessary for this case and the theoretical use of a Kerr-type response would, in all probability, not be appropriate.

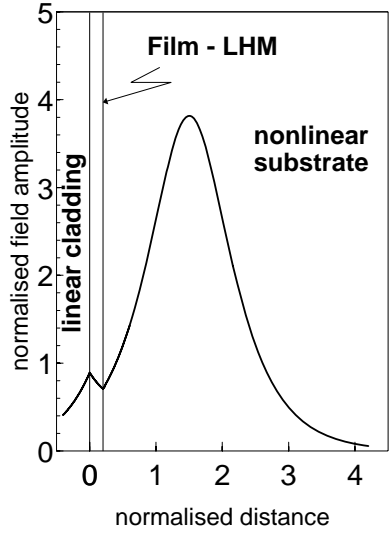

Figure 2 *An example of* (7) *that gives a "bulge" of the electric field amplitude for a planar thin film wave guide with a nonlinear substrate with zero-field permittivity 2.5, and an air cladding. The normalised field is* $\sqrt{\gamma_s/2}E$ *and the normalised distance is* $cd/\omega_p$ *where d is the actual distance in m.*
*The permittivity and permeability of the film are given in* (1).

The choice of a minus sign before the left-hand-side of (10) with plus signs in the numerator and denominator, however, gives a dispersion equation with a linear limit, i.e.

$$\tanh(k_f d) = -\frac{k_f \mu_f \left[\sqrt{k_z^2 - \frac{\omega^2}{c^2}\varepsilon_c} + \sqrt{k_z^2 - \frac{\omega^2}{c^2}\varepsilon_s}\right]}{k_f^2 + \mu_f^2 \sqrt{\left[k_z^2 - \frac{\omega^2}{c^2}\varepsilon_c\right]\left[k_z^2 - \frac{\omega^2}{c^2}\varepsilon_s\right]}} \qquad (11)$$

and it is reasonable to assume that the use of the Kerr-type nonlinearity will give acceptable results. Thus, the effect of a Kerr type nonlinearity on the dispersion of the TE surface waves given by (11) will now be examined.

In the RHM substrate and cladding the group and phase velocities are in the same direction (forward travelling waves). In the LHM film the group and phase velocities are in opposite directions (backward travelling waves). In the planar wave guide there is thus a competition between the forward and backward travelling waves, and the total power flow will be either forwards, or backwards, depending on the sum of the positive flowing power in the substrate and cladding, and the negative flowing power in the film. This gives a characteristic maximum in the dispersion equation for the TE surface waves in linear LHM planar waveguide. Now equation (11) shows that the position of the dispersion curve at which this maximum occurs for a fixed value of, say, the cladding permittivity, is determined by the substrate permittivity. If, then, the substrate is nonlinear, in which case the permittivity depends on the intensity of the electric field, the position of the maximum in the dispersion curve can be controlled by the intensity of the field.

Since the maximum in the dispersion curve indicates the value of the wave number at which the total power flow changes from positive to negative, then there is a possibility that the direction of the power flow can be controlled by the intensity of the electric field.

The dispersion equations (10) have corresponding nonlinear power flows in the substrate and cladding given by

$$P_s = \frac{2}{\gamma_s}\frac{\varepsilon_0 k_z}{2\omega^3}\left[k_s \pm \left(k_s^2 - \frac{\omega^2}{c^2}\frac{\gamma_s}{2}E_0^2\right)^{\frac{1}{2}}\right], \quad P_c = \frac{2}{\gamma_c}\frac{\varepsilon_0 k_z}{2\omega^3}\left[k_c \pm \left(k_c^2 - \frac{\omega^2}{c^2}\frac{\gamma_c}{2}E_d^2\right)^{\frac{1}{2}}\right] \quad (12)$$

where the plus signs are relevant for field profiles that increase away from the ,respective, boundary in the substrate and/or cladding. The minus signs indicate immediate decay away from the boundaries into the substrate and cladding respectively, [a minus sign in both $P_s$ and $P_c$ gives the corresponding linear case as is shown by taking the limits of (12), as $\gamma_{s,c}$ approach zero]. The relationship between the boundary field amplitudes is

$$\left(\mu_f^2 k_s^2 - k_f^2 - \mu_f^2\frac{\omega^2}{c^2}\frac{\gamma_s}{2}E_0^2\right)E_0^2 = \left(\mu_f^2 k_c^2 - k_f^2 - \mu_f^2\frac{\omega^2}{c^2}\frac{\gamma_c}{2}E_d^2\right)E_d^2 \quad (13)$$

A detailed examination of (13) will appear in a later work. Here (13) is used to find the value of $E_d$ in terms of $E_0$ for solution of (12).

Figure 3 shows the shift in the dispersion curve of the TE surface waves due to an increase in intensity and the corresponding increase in substrate permittivity. Figure 4 shows the corresponding positions of the power flow change from positive to negative.

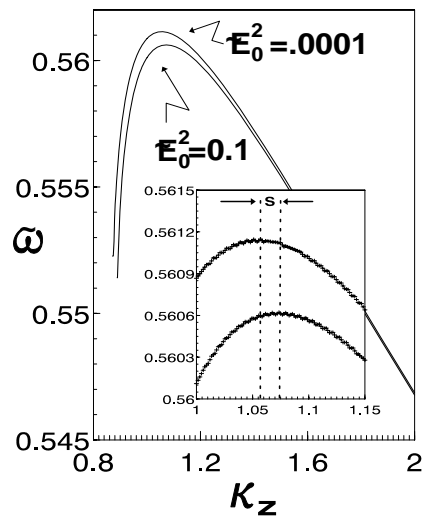

Figure 3. *The shift* (S) *in the position of the maximum in the dispersion curves in the planar LHM waveguide as the intensity of the electric field at the LHM/substrate boundary increases.* $\varepsilon_c = 1$, $\varepsilon_s = 2.5$, *other parameters as in figure* 1.

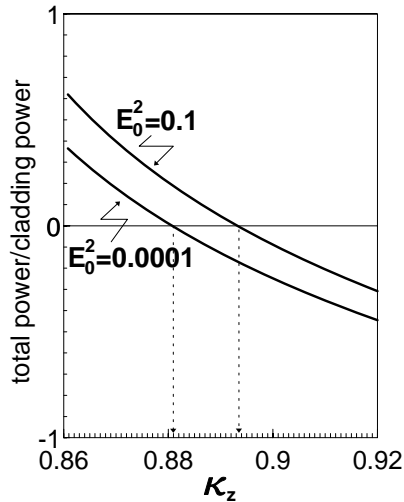

Figure 4. *The shift in the position of the change from positive to negative power corresponding to the shift in the maximum of the dispersion curve in figure 3.*

The field profiles shown in figure 5 demonstrate clearly that the fraction of the field that is carried in the nonlinear substrate is a function of the intensity. As the intensity increases the fraction of the field in the substrate increases thus increasing the power flow in the substrate compared to that flowing in the LHM film. This suggests the possibility of determining the direction of the power flow through the intensity of the electric field.

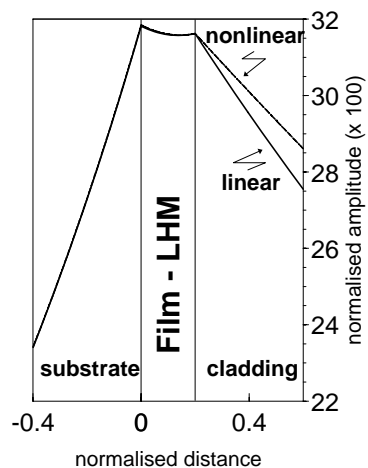

Figure 5. Electric field profile across the linear/LHM/nonlinear planar guide described in the text. The nonlinearity increases the positive power flow in the nonlinear substrate compared to the negative power flow of the LHM film. Parameters as in figure 2.3.

## 3. The FDTD approach

A major disagreement has arisen in the literature [35] that casts doubt upon the phenomenon of negative refraction because, looked at in certain way, it appears to violate causality and requires light to travel instantaneously from one point to another. The source of this disagreement really concerns the difference between a single plane wave and a light beam. The latter can be viewed as being constructed from an infinite sum of plane waves, all with different wave vectors. If it is believed that negative refraction can occur for a single plane wave then it is difficult to see how negative refraction can occur for a causally-driven beam of finite width propagating in a dispersive medium. The resolution to this question ought to be found using a tool that permits the creation of a *causal finite beam* in a natural way and then reveals the dynamical development of it as it encounters a left-handed medium.

Keeping in mind an *optical frequency* application [15,16] of left-handed behaviour, a two-dimensional finite-difference-time-domain (FDTD) computational scheme is just such a tool [36 - 40] and will be used here to observe how a spatial soliton at optical frequencies will behave when it encounters an interface between a right-handed medium and a left-handed medium. The method deploys the standard Yee algorithm [38] and involves numerically solving Maxwell's equations directly. As others have pointed out, it is a method that is free of approximations, is a full-wave vector analysis tool, does not suffer from the constraints of paraxiality and above all permits advanced material modelling, coupled to a variety of source excitations. In the two-dimensional cases explored here ($\partial/\partial z = 0$), TE polarised beam interactions are investigated that require only the three field components $E_z$, $H_x$ and $H_y$, in an *xy*-plane bounded by a standard 20 cell thick perfect matched layer absorbing boundary [36]. The basic geometry for the simulations is presented in figure 6, where four regions are identified but only three different media types are used. An isotropic, lossless LHM slab is introduced but this time it is the Drude model that is adopted. This means that the dispersion introduced by the relative permittivity and relative permeability is now given by

$$\varepsilon(\omega) = 1 - \frac{\omega_{pe}^2}{\omega^2} \quad (14)$$

$$\mu(\omega) = 1 - \frac{\omega_{pm}^2}{\omega^2} \quad (15)$$

Note that, although this model is a little different from the one adopted for section 2, in so far as it does not contain a resonance frequency, it is one that has been driving other FTTD work in the field. It is not self-evident that this model of simultaneous negative permeability and permittivity will suffice for optical nanostructure materials but it is still taken here as working model to demonstrate the points being exposed. The conclusions will be qualitatively the same, whichever model emerges as the final one. The LHM is also assumed to be linear in the model calculation. This means that the spatial soliton is sustained by RHM bounding media that possess a dominant nonlinearity. In any case, at optical frequencies it is assumed that the nonlinearity in the LHM will either be too weak to make an impact, or that the bounding media will be deliberately selected to be dominant. The matching RHM that will be paired with the LHM slab is also assumed to be linear and to be matched in terms of refractive index. This situation does not have to be with material found in Nature: it can also be engineered artificially and be also controlled by frequency. All the partners in this soliton interface arrangement can also have their losses controlled to a minimum. The structure analysed below is based upon the foregoing arguments and although, at first sight appears to be highly stylistic, it should be capable of being engineered.
The nonlinear regions are assumed to be capable of the kind of self-focusing that arises from an unsaturated Kerr nonlinearity. They are right-handed and are assumed to contain a material like type-RN Corning glass [37]. Obviously, high powers would be needed then but the qualitative features that emerge from the simulations reported here are not affected by this qualification. For the right-handed nonlinear medium, the example data for the linear relative permittivity and relative permeability are $\varepsilon = (2.46)^2$, $\mu = 1$, respectively, so the linear refractive index is set to the value $n_0 = 2.46$. The material nonlinearity is assumed to be third-order with a susceptibility coefficient $\chi^{(3)}$. The total nonlinear refractive index is then $n = n_0 + n_2 I$, where $I$ is the beam intensity and, typically, $n_2 = 1.25 \times 10^{-18}$ m$^2$/W is the value set for the computations. In order to minimise reflections at all the interfaces the wave impedances are artificially matched. This is assured by setting the data of the LHM to ensure that the plasma frequencies $\omega B_{pe}$ and $\omega_{pm}$ give $\varepsilon(\omega_c) = -(2.46^2)$, $\mu(\omega_c) = -1$, at the carrier frequency $\omega_c$. This form of impedance matching is not an 'in principle' limitation. Unmatched cases merely have reflections appearing at the interfaces and their presence does not alter the fundamental conclusions being sought here. For this reason, the lens is a matched pair of entirely *linear* materials 'a sort of black box' [11] that has a *linear* matching RHM, that will be a different material, with an *engineered* refractive index to match also that of the bounding Kerr media. The beam will diffract and disperse as it enters the LHM and so it should spread out as it approaches the linear LHM/RHM interface. For a real laboratory experiment this sort of material engineering may be very difficult to stabilise but the points brought out in the simulations reported here will still be in evidence. Naturally, if this type of superlensing [41,42] is being sought then a closely impedance-matched scenario *must* be developed. Finally, the linear right-handed slab attached to the LHM is assumed also to be *dispersionless*. It must be emphasised, once again, that the data is selected not only for computational and visual convenience, to minimise impedance mismatching, but also on the grounds that the structure could be engineered artificially.

It has already been commented for superlensing [11] that maintaining matching is an important issue. Nevertheless, the points to be brought out below will only change quantitatively, not qualitatively, by adopting mismatched indices and introducing loss, about which more will be said later.

An x -directed spatial soliton is assumed as starting point [43] and this is described, in the familiar way, by the standard equation

$$E_z(x=0, y) = \frac{1}{kw}\left(\frac{1}{n_o n_2}\right)^{\frac{1}{2}} \text{sech}\left(\frac{y - y_0}{w}\right) \qquad (16)$$

where $k$ is the wave number, $w$ the characteristic width and $y_0$ is the beam centre. This soliton is *injected* into the RHM at $x = 0$ and with a sinusoidal temporal profile. An interesting point now arises. If the beam is turned on very slowly then its bandwidth [44] can be kept under control and it has been argued that reducing the bandwidth permits a rather easier interpretation of what happens at an interface. It is concluded here that this is not an important point, from the point of view of interpretation, so the turn-on speed (time-dependence) is not optimised to guarantee a small bandwidth. This improves the computational speed but the somewhat larger number of frequency components thereby generated does not impede the interpretation at all.

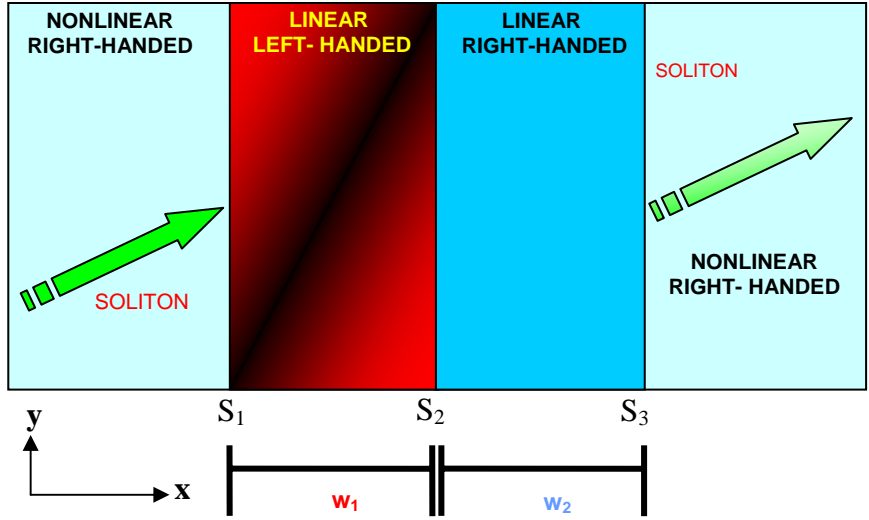

**Figure 6.** Outline of a basic 'soliton lens' comprising a region of LHM of width $w_1$ and a region of linear RHM of width $w_2$ where $S_1$, $S_2$ and $S_3$ denote the locations of the material interfaces

One of the flexibilities of the FDTD method is that the simulation is scaled relative to the wavelength being studied. This means that even though in this case $\lambda = 1.3$ $\mu$m but the results given here are still generic. The simulations use a point separation $\Delta = 0.065$ $\mu$m and a time step $\delta T = 108.408 \times 10^{-18}$ s. The soliton beam width is 0.65 $\mu$m and this means that the effective Rayleigh length is 77.28 points, which is 5.0232 $\mu$m, in real units. The beam width is rather narrower than is normally used when only using the nonlinear Schrödinger equation as a computational tool. This is one of the advantages of the FDTD approach because it is not restricted by the avoidance of a non-paraxial regime. The soliton shown here is one-dimensional, and the diffraction that takes place in the direction normal to the propagation direction is opposed by the self-focusing. Such beams are marginally stable but, because $w = \lambda/2$, the beam is non-paraxial and the two-dimensional counterpart would also be stable and in fact become an *optical needle* [45,46]. It is beyond the scope of the present paper but ultra-narrow beams with very high light intensities are obviously of some technical importance for future applications. This topic will be returned to in later publications but for now the first interaction of narrow beams with LHM will be discussed

## 4.     Numerical Results for the soliton lens

Upon entering the linear, dispersive LHM, a spatial soliton beam will diffract, after experiencing negative (anomalous) refraction. If the bandwidth, created by the turn-on speed of the source is sufficient to access the dispersion significantly then this will also overlay the diffraction. However, it is the ability of the RHM to 'cancel out' [11] the effect of the LHM (basically an anti-refraction process, analogous to antimatter waves) that prompts the present investigation. Of course it is not a new speculation, in general, and has been brilliantly used by Pendry to discuss entirely linear situations [11,12] that can produce aberration-free images. What is new here is the use of narrow spatial solitons of the kind that will be found in optical chip-level circuit designs of the future.  It is important to be able to deliver optical needle-like beams to different parts of the circuits by indulging in a phase compensation offered by RHM/LHM cancellation properties.  The extent to which the soliton is 'repaired' throughout this cancellation experience is an important property to measure, and some correlation work is included later on for that purpose. Naturally, there will be absorption in any real system but this can be 'repaired' by including amplification, or generally using so-called dissipative solitons. Loss compensation is not expected to disturb the generic conclusions reached here, because we are using *sech*-like inputs.  For super-Gaussian inputs, the inclusion of processes like amplification, higher-order nonlinearity, and damping mechanisms other than linear in origin, will be the subject of further investigation. To state again what is expected, an RHM could be paired with an LHM to permit the distortion-free transmission of solitons through an entirely linear region. In reality, this 'cancelling' effect is a direct result of the sudden direction change experienced by the phase velocity of the beam upon transmission through a LHM/RHM interface. Any phase that the beam acquires through the LHM will basically 'run backwards' in the RHM. In the particular instance when the linear regions are of equal width, the phase evolution in the LHM is exactly reversed by passage through the RHM. This is expected but the question is not only whether the FDTD approach shows up negative refraction in a perfectly natural way but whether this anomalous refraction can be undone by the RHM for all angles of incidence and under what conditions will solitons fail to be reformed. Nobody has addressed these questions yet for nonlinear beams.

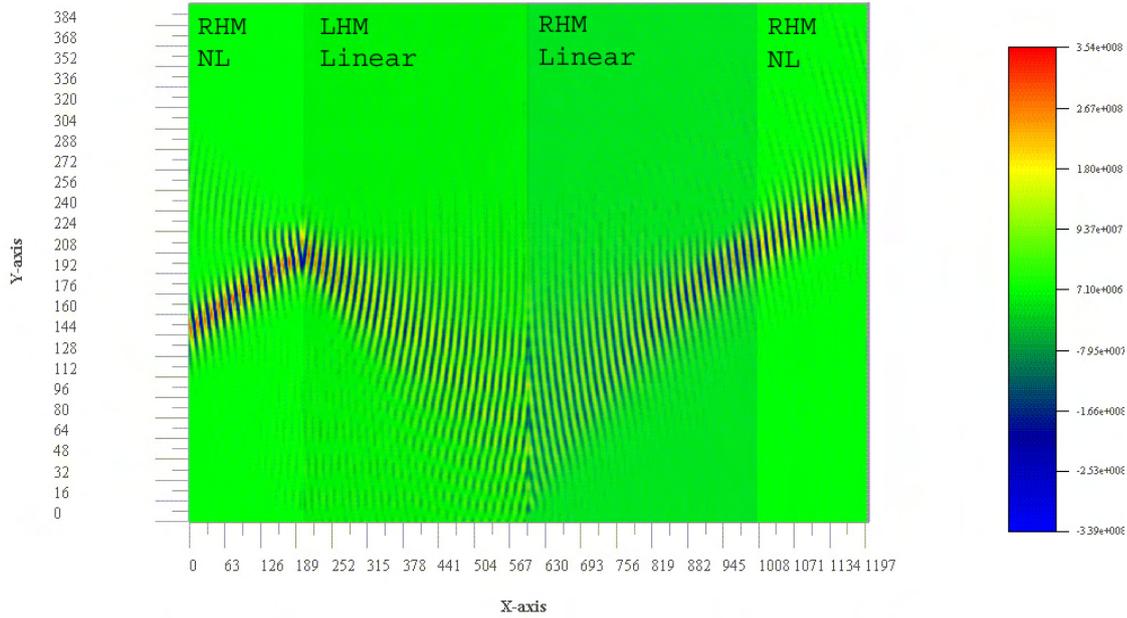

**Figure 7** Interaction of a soliton with an incident angle of approximately 20º with a soliton lens possessing equal widths of RHM and LHM. This is an electric field plot at time $T = 11000 \, dT$.

Figure 7 first of all shows the LHM 'undoing' the spatial soliton beam. This process is mainly diffraction although there is some evidence of dispersion attacking the finite bandwidth. The main point is that the FDTD causal calculation shows clearly the negative diffraction, as indeed others have observed in their simulations. The beam is narrow and nonparaxial and an angle of incidence of $20^0$ is accurately sustained by the calculation. The role played by the LHM is completely reversed when it enters the right-handed region and the visual evidence is such that it looks as if the spatial soliton has recovered its identity in the emergent nonlinear medium. In essence, the soliton loses its identity for a certain distance but it is retrieved in a way that makes it seem that the soliton never interacted at all with the two linear slabs. This will a very useful feature in all-optical chip design and the behaviour can be explained qualitatively by using the ideas of complementary media so beautifully outlined by Pendry and his collaborators. For future design purposes, however, a quantitative method is put forward here that is based upon correlation techniques.

The beam profile at an *x*-position $S_1 + a_x$, corresponding to a point $a_x$ *inside* the LHM can be recorded. If the soliton lens is operating in a way that relies upon negative refraction, then it is expected that this profile will be almost identical to the profile of the beam at the position $x = S_1 + S_2 + S_3 + a_x$. Two separate simulations can then be run in tandem; the first will rely upon 1400 × 500 points, with the LHM/RHM lens lying between $x = 201$ and 1000, while the second (reference) will rely upon 600 × 500 points comprising solely the bulk nonlinear medium. Profiles can then be taken at one point before the lens then from 1 to 200 points after it, permitting a comparison using a circular cross-correlation technique [47]. In order to avoid complications caused by the fast variations, the profile correlations should be conducted using time-averaged electric field amplitudes. To study the effects of the lens on the structure of the emerging soliton two factors can be investigated  First, the angle of incidence of the input beam can be varied from 0º to 35º , for fixed widths $w_1 = w_2$. Secondly, for a fixed angle of 20º (arbitrarily chosen based on the simulation capacity) the ratio $w_1/w_2$ can be varied, by just varying $S_2$ : this keeps the extremes of the lens fixed.

*(a)    Soliton angle of incidence*

For the perfect design where $w_1 = w_2$, unsurprisingly, the soliton is almost perfectly reconstructed for a wide range of incident angles. In fact, it is practically independent of the incident angle and experiences virtually no lateral *y*-shift as a result of any interaction within the lens. Figure 8 gives a typical result, in which the angle of incidence is equal to 20º. It compares the extracted *y*-profiles and displays the circular correlation plot for a randomly chosen distance from the lens. The result portrayed here applies to all the other cases where the position of one profile relative to the other (or the shift) is always zero with a correlation factor or approximately 99%. Although the correlation is very high it is not exactly one. This result can be explained due to the fact complete impedance matching is impossible, because of the role of the nonlinearity.  Cross-correlation normalises the data first, so it is independent of amplitude, and is based purely on the aspect of the data, not the actual numerical values. Hence any outcome cannot be affected by a reduction in amplitude incurred in the lens system. A correlation factor of 1 indicates that the normalised profiles can be exactly superimposed whereas 0 indicates that there is no correlation. Purely for visualisation, in order that the two profiles can be clearly seen, the profile taken from the simulation including the lens is down-shifted by 0.5.

With the time averaging, the clock was started at the same time as the simulation i.e. at $T = 0$. The FDTD gives a snap-shot at an instant in time including the fast variation. This fast variation can be effectively removed by taking the time-average of the amplitude |*E*|. However, since the averaging is started earlier, when the beam is switched on, the regions the beam first encounters will have a higher average than those which it only reaches in the final few time steps. In other words, the regions closer to the source have received more energy during the total simulation time than those further down the beam path. Hence, even with *no losses* the time-averaged electric field amplitude, in this case, will always show a decrease from a maximum at the source line to a minimum at the far edge of the simulation region.

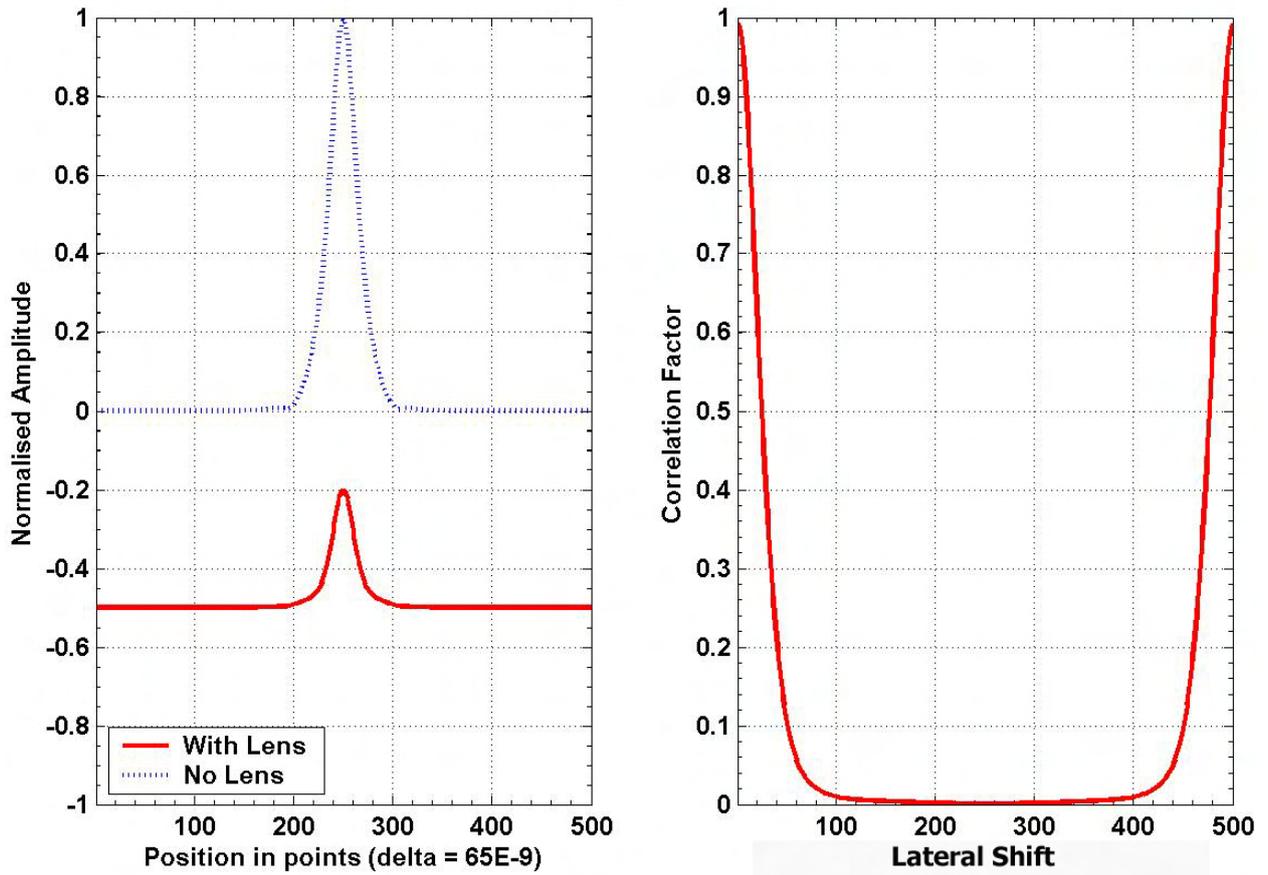

**Figure 8a:** Extracted *y*-profiles taken from equivalent *x* positions (taking into account the lens width of 800 points) both normalised relative to the peak time averaged amplitude of the reference beam in the bulk nonlinear media – dotted line. Note that the reduced amplitude of the profile taken with the soliton lens present is reduced mainly as an artefact of the time averaging and *not* energy loss.

**Figure 8b:** Circular cross-correlation of the two extracted profiles depicting the correlation factor as a function of the position in the *y*-direction of one profile relative to the other. By the nature of the circular aspect of this technique the profiles are cyclically correlated, i.e. the points that 'fall off' the edge are wrapped around.

Importantly, the correlation factor remains constant with distance, as measured from the edge of the lens. This means that the emerging beam possesses the same form as that of the incident spatial soliton, for all the times addressed in these simulations. The idea, as constantly emphasised by Pendry , is that the influence of the LHM and RHM effectively cancel each other out. This certainly seems to be holding up very well when applied to incident solitons entering the lens at a wide range of angles, but the widths do need to be *matched.* As commented for the perfect lens scenario, in practice this will require careful control of the bandwidth when the soliton is launched. However, it is interesting to note that the solitons do have to pass first into a linear *dispersive* media (the LHM) before entering a linear non-dispersive medium. This makes the beam reconstruction more interesting. A question now arises as to the effects of the width of the lens blocks on the emerging soliton and this will now be discussed.

*(b)    Ratio of slab widths within the soliton lens*

The width ratios can be explored in pairs around the impedance matched condition, and a selection has been made that examines the outcomes of using rations ranging from $w_1 = 3 \times w_2$ down to $w_1 =$ $w_2/3$. The beam is expected to undergo a lateral translation in the y- direction, provided everything is linear but this is the first investigation that studies the lateral shift at the interface of a nonlinear-left handed material. What is particularly interesting is that for a given ratio. $w_1 = w_2 \times R$ the shift is equal, yet opposite, to the shift for the reciprocal ratio $w_1 = w_2/R$ ,at least for the outcomes shown in figure 9.

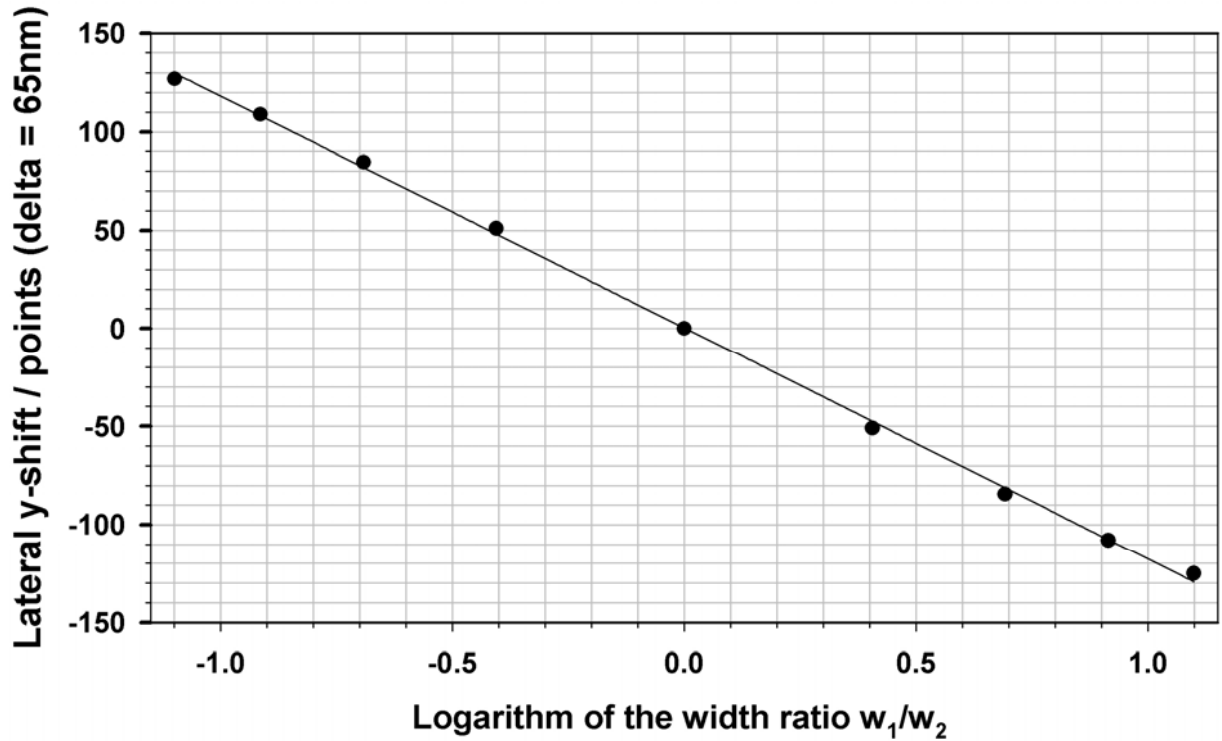

**Figure 9.** Lateral y-shift experienced by a spatial soliton beam as a function of the natural logarithm of the LHM/RHM width ratio

Of equal interest is the effect on the correlation factor immediately after emerging from the lens because this factor decreases as the ratio moves away from the optimal value of 1. The variation of the subsequent beam behaviour depends on whether the LHM is thinner, or thicker, than the RHM. If it is thinner, the diffraction the beam experiences in the LHM is reversed well within the RHM passing through a focus and then subsequently undergoing further diffraction. As a result, the emerging beam is not capable of reforming a soliton; thus the soliton is destroyed. Normal diffraction occurs and as a result the correlation steadily decreases with distance from the lens, as shown in figure 10.

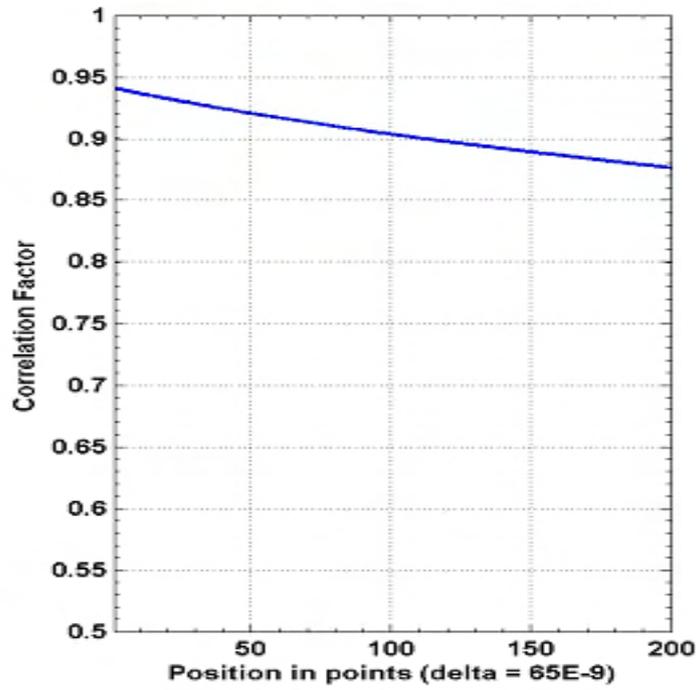

**Figure 10.** Correlation factor with distance from the lens edge for a width ratio of 2/3

In the latter case, the diffraction in the LHM is not fully reversed in the linear RHM. Instead, it undergoes completion and therefore approaches the focus within the nonlinear RHM medium. This can be seen by a steady increase in the correlation factor, up to a maximum. However, during this increase the self-focusing and anomalous diffraction are working in tandem and appear to compress the beam such that at focus it would be unstable. When the diffraction switches in character, the correlation factor dips slightly as the soliton re-stabilises, as seen in figure 11.

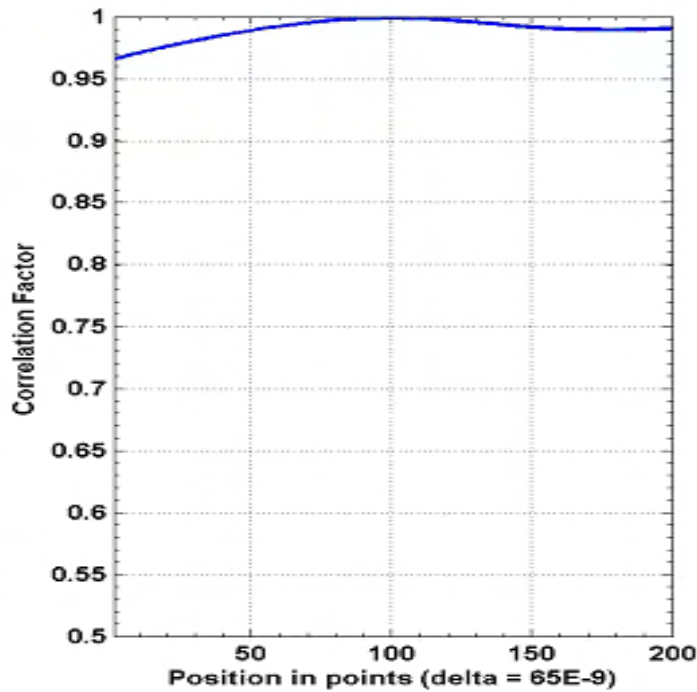

**Figure 11:** Correlation factor with distance from the lens edge for a width ratio of 3/2

## 6. Conclusions

This paper discusses the behaviour of nonlinear waves in a left-handed medium. The opening discussion highlights the fact that a left-handed medium is a member of quite a broad class of materials that exhibit negative refraction as a major characteristic. It is pointed out that Schuster [20] who was Professor of Physics in Manchester was the first to discuss this possibility in 1904 and that he cited the fact that within the absorption band of, for example, sodium vapour, will propagate as a backward wave. Media with an effective negative permittivity, and an effective negative permeability, are a class of isotropic materials that can cause negative refraction. They can now be manufactured to operate at frequencies ranging from microwave to optical, and this development is ushering in a new era of metamaterials, especially in the optical range which deploy nanostructured materials. Such materials are referred to here as 'left-handed', even though a variety of names for them is evident from the literature. In anticipation of a demand for highly structured, integrated, programmable, practical waveguides, this paper addresses the joint role of left-handedness and nonlinearity. First, a planar guide is investigated, in which the waveguide is a slab consisting of a left-handed medium sandwiched between a substrate and cladding that are simple dielectrics. It is shown that the uses of left-handedness in a waveguide produces new properties that enable a detailed control of the power flow, even at substantially low mode intensities. After this discussion a comprehensive finite-difference-time-domain (FDTD) analysis is presented that concentrates upon spatial soliton behaviour. These spatial solitons have a width that is a fraction of the wavelength of light and, therefore, are 'optical needles'. They are highly non-paraxial and can be incident at any angle upon a right-handed-left-handed planar interface. This is a huge merit of the FDTD approach. An interesting soliton-lens arrangement is investigated that lends itself to a novel cancellation effect. The cancellation could be subject to tolerances, such as angle of incidence choice or width ratios so a new approach is to introduce a circular cross-correlation technique that gives a quantitative measure of the robustness of the now famous RHM/LHM phase compensation. It is commented that in an experimental design, this correlation work could be considerably expanded to discuss the range of permitted parameters that will lend themselves to all-optical chip design work. The desire to use left-handed material is not only to achieve negative refraction but also to create new waveguide and interface designs that have plasmons as a natural ingredient. Finally, it is believed that the recent nanostructure work [15,16] is pointing to the way things are going to develop, in the next few years.